\documentclass[preprint,amsmath,amssymb,aps,tightenlines]{revtex4}

\usepackage{graphicx}
\usepackage{subfigure}

\def\beq{\begin{equation}}
\def\eeq{\end{equation}}
\def\barr#1{\begin{array}{#1}}
\def\earr{\end{array}}
\def\beqar{\begin{eqnarray}}
\def\eeqar{\end{eqnarray}}
\def\beqars{\begin{eqnarray*}}
\def\eeqars{\end{eqnarray*}}
\def\bitem{\begin{itemize}}
\def\eitem{\end{itemize}}

\def\bc{\begin{center}}
\def\ec{\end{center}}
\def\beq{\begin{equation}}
\def\eeq{\end{equation}}
\def\bea{\begin{eqnarray}}
\def\eea{\end{eqnarray}}
\def\bit{\begin{itemize}}
\def\eit{\end{itemize}}
\def\ben{\begin{enumerate}}
\def\een{\end{enumerate}}
\def\ba{\begin{array}}
\def\ea{\end{array}}
\def\bc{\begin{center}}
\def\ec{\end{center}}

\def\dReg(#1,#2){\SetOffset(#1,#2)\BCirc(0,0){4}
    \Line(2.8,2.8)(-2.8,-2.8)
    \Line(2.8,-2.8)(-2.8,2.8)\SetOffset(0,0)}

\newenvironment{rowvec2}{\left(\begin{array}{cc}}{\end{array}\right)}
\newenvironment{colvec}{\left(\begin{array}{c}}{\end{array}\right)}
\def\bcol{\begin{colvec}}
\def\ecol{\end{colvec}}
\def\brow2{\begin{rowvec2}}
\def\erow2{\end{rowvec2}}

\begin{document}
%\today
%\draft
\preprint{\begin{tabular}{l}
\hbox to\hsize{June, 2002 \hfill DESY 02-85}\\[0mm]
\hbox to\hsize{hep-ph/0206297 \hfill KAIST-TH-2002/15 }\\[5mm] \end{tabular} }

%\preprint{KAIST-TH-xx,\\ hep-ph/0202xxx}

\bigskip

\title{$B^0 - \overline{B^0}$ mixing, $B\rightarrow J/\psi K_s$ and 
$B\rightarrow X_d ~\gamma$ \\ in general MSSM }

\author{P. Ko}
\address{Department of Physics, KAIST \\ Daejeon 305-701, Korea}

\author{G. Kramer}
\address{ II. Institut f\"{u}r Theoretische Physik, 
Universit\"{a}t Hamburg \\ D-22761 Hamburg, Germany}

\author{Jae-hyeon\ Park}
\address{Department of Physics, KAIST \\ Daejeon 305-701, Korea}
\date{\today}

\bigskip
\bigskip
\bigskip

\vspace{0.5in}
\begin{abstract}
% abstract
We consider the gluino-mediated SUSY contributions to $B^0 - \overline{B^0}$ 
mixing, $B\rightarrow J/\psi K_s$ and $B\rightarrow X_d \gamma$ in the mass
insertion approximation.
We find the $(LL)$ mixing parameter can be as large as 
$| (\delta_{13}^d)_{LL} | \lesssim 2 \times 10^{-1}$, 
but the $(LR)$ mixing is strongly constrained by the $B\rightarrow X_d \gamma$ 
branching ratio and we find $| (\delta_{13}^d)_{LR} | \lesssim 10^{-2}$.
The implications for the direct CP asymmetry in $B\rightarrow X_d \gamma$ 
and the dilepton charge asymmetry ($A_{ll}$) are also discussed, where 
substantial deviations from the standard model (SM) predictions are possible.
\end{abstract}

%\pacs{}

\maketitle

\newpage
%\narrowtext

%  body

%%%%%%%%%%%%%%%%%%%%%%%%%%%%%%%%%%%%%%%%%%%%%%%%%%%%%%%%%%%%%%%%%%%%%%%%%%%
\section{Introduction}
%%%%%%%%%%%%%%%%%%%%%%%%%%%%%%%%%%%%%%%%%%%%%%%%%%%%%%%%%%%%%%%%%%%%%%%%%%%

Recent observations of large CP violation in $B\rightarrow J/\psi K_s$
\cite{exp:sin2beta,masiero2002} giving 
\begin{equation}
\sin 2 \beta = ( 0.79 \pm 0.10 )
\end{equation}
confirm the SM prediction and begin to put a strong constraint on new physics
contributions to $B^0 - \overline{B^0}$ mixing and $B\rightarrow J/\psi K_s$,
when combined with $\Delta m_{B_d} = (0.472 \pm 0.017 )~{\rm ps}^{-1}$ 
\cite{pdg}. Since the decay $B\rightarrow J/\psi K_s$ is dominated by the 
tree level SM process $b\rightarrow c \bar{c} s$, we expect the new physics 
contribution may affect significantly only the $B^0 - \overline{B^0}$ mixing
and not the decay $B\rightarrow J/\psi K_s$. A model independent study of
$B^0 - \overline{B^0}$ mixing has been discussed recently by Laplace et al.
\cite{nir}. However, in the presence of new physics contributions to 
$B^0 - \overline{B^0}$ mixing, the same new physics would generically affect 
the $B\rightarrow X_d \gamma$ process. The relation between the new physics 
effects on the $B^0 - \overline{B^0}$ mixing and $B\rightarrow X_d \gamma$ 
is in principle independent of each other and one may adopt a model 
independent analysis based on effective lagrangian with dimension 5 or 6
operators (for example, see Ref.~\cite{kkl} for the model independent 
study of the correlation between $B\rightarrow X_s \gamma$ and $B\rightarrow
X_s l^+ l^-$. The second paper in Ref.~\cite{kkl} includes a new CP 
violating phase in the $C_{7\gamma}$ Wilson coefficient.). 
This approach would introduce 4 new independent complex 
parameters: two in the $B^0 - \overline{B^0}$ mixing, and two in the 
$B\rightarrow X_d \gamma$. Being with too many independent parameters, 
one would not be able to get definite predictions based on this approach.

In this work, we do not attempt a completely model independent study with 
too many new independent parameters. Instead, we  consider 
$B^0 -  \overline{B^0}$ mixing, $B\rightarrow J/\psi K_s$ and 
$B_d \rightarrow X_d \gamma$ in general SUSY models where flavor and CP 
violation due to the gluino mediation can be important. 
The chargino-stop and the charged Higgs-top loop contributions are 
parametrically suppressed relative to the gluino contributions, and thus 
ignored following Ref.~\cite{antichi}.
(See however Refs.~\cite{ali1,kane} for including such effects. Another popular
approach which is orthogonal to our approach is the minimal flavor violation 
model, which is discussed in Refs.~\cite{mfv} in the context of $B$ physics.)
We use the mass insertion approximation (MIA) for this purpose.
Comprehensive work has been done for the first two observables in the MIA
considering $\Delta m_{B_d}$ and $\sin2\beta$ constraints only (see Ref.~
\cite{masiero2002}  for the most recent studies with such an approach). 
In our work, 
we also include the dilepton charge asymmetry $A_{ll}$ and the 
$B_d \rightarrow X_d \gamma$ branching ratio constraint extracted from the 
recent experimental %errors in the $B\rightarrow X_s \gamma$ measurement, 
upper limit on the $B\rightarrow \rho\gamma$ branching ratio \cite{b2rho}
\[
B( B \rightarrow \rho \gamma ) < 2.3 \times 10^{-6}, 
\] and rederive
the upper limits on the $(\delta_{13}^d )_{LL}$ and $(\delta_{13}^d )_{LR}$ 
mixing parameters assuming that only one of these gives a dominant SUSY 
contribution in addition to the standard model (SM) contribution. 
In addition we study the direct CP asymmetry in $B_d \rightarrow X_d \gamma$
on the basis of our result for the SUSY contribution, and 
discuss how much deviations from the SM predictions are expected. 
Although we confine ourselves here to the gluino-mediated SUSY constributions 
only, our strategy can be extended to any new physics scenario
with a substantial constribution to $B^0 - \overline{B^0}$ mixing and 
$B\rightarrow X_d \gamma$. 

%%%%%%%%%%%%%%%%%%%%%%%%%%%%%%%%%%%%%%%%%%%%%%%%%%%%%%%%%%%%%%%%%%%%%%%%%%%
\section{Effective Hamiltonians for $B^0 -\overline{B^0}$ mixing 
and $B\rightarrow X_d \gamma$ }
%%%%%%%%%%%%%%%%%%%%%%%%%%%%%%%%%%%%%%%%%%%%%%%%%%%%%%%%%%%%%%%%%%%%%%%%%%%

\subsection{Effective Hamiltonian for $B^0 -\overline{B^0}$ mixing }

The most general effective Hamiltonian for $B^0 -\overline{B^0}$ mixing
($\Delta B = 2$) can be written in the following form \cite{masiero2002}:
\begin{equation}
\label{eq:eh}
H_{\rm eff}^{\Delta B = 2} = \sum_{i=1}^5 C_i Q_i + 
\sum_{i=1}^3 \tilde{C}_i \tilde{Q}_i, 
\end{equation}
where the operators $Q_i$'s are defined as
\begin{eqnarray}
Q_1 & = &  \bar{d}_L^{\alpha} \gamma_\mu b_L^{\alpha}~
           \bar{d}_L^{\beta}  \gamma^\mu b_L^{\beta}\,
\nonumber  \\
Q_2 & = &  \bar{d}_R^{\alpha} b_L^{\alpha}~\bar{d}_R^{\beta} b_L^{\beta}\, 
\nonumber  \\
Q_3 & = &  \bar{d}_R^{\alpha} b_L^{\beta} ~\bar{d}_R^{\beta} b_L^{\alpha}\, 
\nonumber  \\
Q_4 & = &  \bar{d}_R^{\alpha} b_L^{\alpha}~\bar{d}_L^{\beta} b_R^{\beta}\, 
\nonumber  \\
Q_5 & = &  \bar{d}_R^{\alpha} b_L^{\beta} ~\bar{d}_L^{\beta} b_R^{\alpha}\,
\label{eq:db2ops}
\end{eqnarray}
and the operators $\tilde{Q}_i$ are obtained from $Q_i$'s by the exchange of 
$L\leftrightarrow R$. $\alpha, \beta$ are color indices, and $q_{L,R} \equiv
(1\mp \gamma_5) q /2$. The Wilson coefficients $C_i$'s receive contributions
from both the SM and the SUSY loops: 
$C_i \equiv C_i^{\rm SM} + C_i^{\rm SUSY}$. 

In the SM, the $t-W$ box diagram generates only contribution to the 
operator $Q_1$, and the corresponding Wilson coefficient $C_1^{\rm SM}$ 
at the $m_t$ scale is given by \cite{db2wilsonSMmW}
\begin{equation}
C_1^{\rm SM} (m_t) = 
\frac{G_F^2}{4 \pi^2} M_W^2 (V_{td}^* V_{tb})^2 S_0(x_t) ,
\end{equation}
where
\begin{equation}
  S_0(x_t) = \frac{4 x_t - 11 x_t^2 + x_t^3}{4 (1 - x_t)^2} 
  - \frac{3 x_t^3 \ln x_t}{2 (1 - x_t)^3} ,
\end{equation}
with $x_t \equiv m_t^2 / m_W^2$.
Performing the RG evolution down to $m_b$ scale 
incorporating the NLO QCD corrections \cite{Buras:1990fn}, 
we get $C_1^{\rm SM}$ at $m_b$:
\begin{equation}
  C_1^{\rm SM} (m_b) = 
  \frac{G_F^2}{4 \pi^2} M_W^2 (V_{td}^* V_{tb})^2 \eta_{2B} S_0(x_t)
  [ \alpha_s (m_b) ]^{-6/23} 
  \left[ 
    1 + \frac{\alpha_s(m_b)}{4 \pi} J_5
  \right] .
\end{equation}
The definition of $J_5$ can be found in Ref.~\cite{Buchalla:1995vs},
and we use the value of the QCD correction factor $\eta_{2B} = 0.551$
therein.
Evaluating the matrix element of $Q_1$,
we set the bag parameter $B_1(m_b)$ in the $\rm \overline{MS}(NDR)$ 
  scheme to 0.87 \cite{Becirevic:2001xt},
  which is numerically equal to the value in the RI-MOM scheme in
  Eqs.~(\ref{eq:bag}).

If the deviation of the squark mass matrix from universality is small,
the SUSY contribution from the gluino-squark loop is best studied in the 
mass insertion approximation,
which renders the flavor structures of the processes manifest.
Flavor violations in the squark sector are parameterized by
sizes of the off-diagonal mass matrix elements relative to
the average squared squark mass,
\begin{equation}
(\delta^d_{ij})_{AB} \equiv (\tilde{m}^d_{ij})_{AB} / \tilde{m}^2 ,
\end{equation}
where $i$ and $j$ are family indices and $A$ and $B$ are chiralities, 
$L$ or $R$.
The mass matrix is understood to be in the super-CKM basis
so that the quark-squark-gluino interaction vertex preserves
flavor.
This method is applicable to a model-independent study
of flavor and/or CP violation in the squark sector
when the series expansion in terms of $(\delta^d_{ij})_{AB}$
is meaningful.
In the presence of general (but small) flavor mixings in the
down-type squark mass matrix, the squark-gluino
box diagrams give the Wilson coefficients \cite{antichi},
\begin{eqnarray}
C_1^{\rm SUSY} & = & 
- {\alpha_s^2 \over 216 \tilde{m}^2}~\left( 24 x f_6 (x) + 66 
\tilde{f}_6 (x)\right) ~\left( \delta_{13}^d \right)_{LL}^2 \,
\nonumber  \\
C_2^{\rm SUSY} & = & - {\alpha_s^2 \over 216 \tilde{m}^2}~204 x f_6 (x) 
~\left( \delta_{13}^d \right)_{RL}^2 \,
\nonumber  \\
C_3^{\rm SUSY} & = & {\alpha_s^2 \over 216 \tilde{m}^2}~36 x f_6 (x) 
~\left( \delta_{13}^d \right)_{RL}^2 \,
\nonumber  \\
C_4^{\rm SUSY} & = & - {\alpha_s^2 \over 216 \tilde{m}^2}~\left[~ 
\left( 504 x f_6 (x) - 72 \tilde{f}_6 (x) \right)~
\left( \delta_{13}^d \right)_{LL} \left( \delta_{13}^d \right)_{RR} \right. 
\nonumber  \\
& & \left.  - 132 \tilde{f}_6 (x) ~
\left( \delta_{13}^d \right)_{LR} \left( \delta_{13}^d \right)_{RL} ~\right]\,
\nonumber  \\
C_5^{\rm SUSY} & = & - {\alpha_s^2 \over 216 \tilde{m}^2}~\left[~ 
\left( 24 x f_6 (x) + 120 \tilde{f}_6 (x) \right)~
\left( \delta_{13}^d \right)_{LL} \left( \delta_{13}^d \right)_{RR} \right. 
\nonumber  \\
& & \left. - 180 \tilde{f}_6 (x) ~
\left( \delta_{13}^d \right)_{LR} \left( \delta_{13}^d \right)_{RL} ~\right] .
\label{eq:db2wilsonSUSY}
\end{eqnarray}
The other Wilson coefficients $\tilde{C}_{i=1,2,3}^{\rm SUSY}$'s are 
obtained from $C_{i=1,2,3}^{\rm SUSY}$ by exchange of $L\leftrightarrow R$.
The loop functions $f_6 (x)$ and $\tilde{f}_6 (x)$,
evaluated in terms of $x \equiv m_{\tilde{g}}^2 / \tilde{m}^2$,
are given by
\begin{eqnarray}
f_6 (x) & = & {6 ( 1 + 3 x ) \ln x + x^3 - 9 x^2 - 9 x + 17 \over 
               6 ( x - 1 )^5 } ,
\nonumber \\
\tilde{f}_6 (x) & = & {6 x ( 1+x)\ln x - x^3 - 9 x^2 + 9 x + 1 \over
               3 ( x - 1 )^5 } .
\end{eqnarray}
These Wilson coefficients
are calculated at $\mu \sim m_{\tilde{g}} \sim \tilde{m}$,
and evolved down to the $m_b$ scale.
A complete NLO RG evolution formula of these Wilson coefficients
is available in Ref.~\cite{masiero2002}.
The initial condition (\ref{eq:db2wilsonSUSY}) is at LO in $\alpha_s$, but 
it would be no problem  to include the NLO correction. For this we use 
\begin{equation}
\label{eq:magic1}
C_r(m_b^{pole})=\sum_i \sum_s
             \left(b^{(r,s)}_i + \eta \,c^{(r,s)}_i\right)
             \eta^{a_i} \,C_s(M_S),
\end{equation}
where the SUSY scale is defined by $M_S = (\tilde{m} +
m_{\tilde{g}})/2$, and $\eta = \alpha_s(M_S) / \alpha_s(m_t)$.  
The list of `magic numbers' $a_i$, $b^{(r,s)}_i$, and $c^{(r,s)}_i$,
in the RI-MOM scheme, can be
found in Ref.~\cite{masiero2002}.
% \begin{equation}
% \label{eq:magic2}
% \begin{array}{l l}
% a_i=(0.286, -0.692, 0.787, -1.143, 0.143)& \\
% & \\
% b^{(11)}_i=(0.865, 0, 0, 0, 0),&
% c^{(11)}_i=(-0.017,0,0,0,0),\\
% b^{(22)}_i=(0,1.879,0.012,0,0),&
% c^{(22)}_i=(0,-0.18,-0.003,0,0),\\
% b^{(23)}_i=(0,-0.493,0.18,0,0),&
% c^{(23)}_i=(0,-0.014,0.008,0,0),\\
% b^{(32)}_i=(0,-0.044,0.035,0,0),&
% c^{(32)}_i=(0,0.005,-0.012,0,0),\\
% b^{(33)}_i=(0,0.011,0.54,0,0),&
% c^{(33)}_i=(0,0.000,0.028,0,0),\\
% b^{(44)}_i=(0,0,0,2.87,0),&
% c^{(44)}_i=(0,0,0,-0.48,0.005),\\
% b^{(45)}_i=(0,0,0,0.961,-0.22),&
% c^{(45)}_i=(0,0,0,-0.25,-0.006),\\
% b^{(54)}_i=(0,0,0,0.09,0),&
% c^{(54)}_i=(0,0,0,-0.013,-0.016),\\
% b^{(55)}_i=(0,0,0,0.029,0.863),&
% c^{(55)}_i=(0,0,0,-0.007,0.019),\\
% \end{array}
% \end{equation}
RG running of $\tilde{C}_{1-3}$ is done in the same way as for $C_{1-3}$.

% [I'M NOT SURE IF THIS IS ABSOLUTELY NESSECARI ! WE MAY USE THE OLD
% VACUUM INSERTION APPROXIMATIONS ! BUT I INVITE YOUR SUGGESTIONS ]
Each matrix element of the $\Delta B = 2$ operators in (\ref{eq:db2ops})
is taken to be
a product of its value in vacuum insertion approximation and
the corresponding bag parameter:
\begin{eqnarray}
\langle B_d | Q_1 (\mu) | \overline{B^0} \rangle & = &
{2\over 3}~m_{B_d}^2 f_{B_d}^2 B_1 ( \mu ),
\nonumber  \\
\langle B_d | Q_2 (\mu) | \overline{B^0} \rangle & = &
-{5\over 12}~\left( { m_{B_d} \over m_b (\mu) + m_d (\mu) } \right)^2~
m_{B_d}^2 f_{B_d}^2 B_2 ( \mu ),
\nonumber  \\
\langle B_d | Q_3 (\mu) | \overline{B^0} \rangle & = &
{1\over 12}~\left( { m_{B_d} \over m_b (\mu) + m_d (\mu) } \right)^2~
m_{B_d}^2 f_{B_d}^2 B_1 ( \mu ),
\label{eq:db2me}
%\nonumber  
\\
\langle B_d | Q_4 (\mu) | \overline{B^0} \rangle & = &
{1\over 2}~\left( { m_{B_d} \over m_b (\mu) + m_d (\mu) } \right)^2~
m_{B_d}^2 f_{B_d}^2 B_1 ( \mu ),
\nonumber  \\
\langle B_d | Q_5 (\mu) | \overline{B^0} \rangle & = &
{1\over 6}~\left( { m_{B_d} \over m_b (\mu) + m_d (\mu) } \right)^2~
m_{B_d}^2 f_{B_d}^2 B_1 ( \mu ).
\nonumber
\end{eqnarray}
Here we use the lattice improved calculations for the bag
parameters in the RI-MOM scheme \cite{Becirevic:2001xt}:
\begin{eqnarray}
B_1 ( m_b ) = 0.87(4)^{+5}_{-4}, && B_2 ( m_b ) = 0.82(3)(4),
\nonumber \\
B_3 ( m_b ) = 1.02(6)(9),        && B_4 ( m_b ) = 1.16(3)^{+5}_{-7},
\nonumber \\
B_5 ( m_b ) = 1.91(4)^{+22}_{-7}.
\label{eq:bag}
\end{eqnarray}
%
% We don't scan over B_i.
%
In addition, we use the following running quark masses in the RI-MOM scheme :
\begin{equation}
m_b ( m_b ) = 4.6 ~{\rm GeV},~~~~ m_d ( m_b ) = 5.4~{\rm MeV}.
\end{equation} 
The bottom quark mass is obtained from the $\overline{MS}$ mass 
$m_b^{\overline{MS}} ( m_b^{\overline{MS}} ) = 4.23$ GeV. 
For the $B_d$ meson decay constant, we assume $f_{B_d} = 200 \pm 30$ MeV.

The above $\Delta B=2$ effective Hamiltonian will contribute to $\Delta m_B$,
the dilepton charge asymmetry and the time dependent CP asymmetry in the decay
$B\rightarrow J/\psi K_s$ via the phase of the $B^0 - \overline{B^0}$ mixing.
Defining the mixing matrix element by
\begin{equation}
M_{12} (B^0) \equiv {1\over 2 m_B}~\langle B^0 | H_{\rm eff}^{\Delta B = 2}
| \overline{B^0} \rangle \,  
\end{equation}
one has $\Delta m_{B_d} = 2 | M_{12} (B_d^0) |$. This quantity is 
dominated by the short distance contributions, unlike the $\Delta m_K$ for 
which long distance contributions would be significant. Therefore the data 
on $\Delta m_{B_d}^{\rm exp}$ will constrain the modulus of $M_{12} (B_d^0)$.
On the other hand, the phase of the $B^0 - \overline{B^0}$ mixing amplitude 
$M_{12} (B^0) \equiv \exp( 2 i \beta^{'} ) ~| M_{12} (B^0) |$
appears in the time dependent asymmetry : 
\begin{equation}
A_{\rm CP}^{\rm mix} ( B^0 \rightarrow J/\psi K_s )= \sin 2 \beta^{'}
~\sin \Delta m_{B_d}t .
\end{equation}
Since there may be large new physics (SUSY in this work) contributions to
both $K^0 - \overline{K^0}$ and $B^0 - \overline{B^0}$ mixings, the CKM
fit may change accordingly. Only those constraints that come from semileptonic
processes may be used, since these will be dominated by the SM contributions
at tree level (unless one considers R-parity violation). Therefore the angle 
$\beta^{'}$ need not be the same as the SM angle $\beta (= \phi_1 )$, and 
the angle $\gamma (= \phi_3)$ should be considered as a free parameter in 
the full range from 0 to $2 \pi$ in principle. This strategy was also adopted 
in some earlier work \cite{ko2,masiero2002}. 

Finally, the dilepton charge asymmetry $A_{ll}$ 
%$\propto {\rm Re}( \epsilon_B )$ 
is also determined by $M_{12} (B^0)$, albeit a possible long distance 
contribution to $\Gamma_{SM} ( B^0 )$. Defining the mass eigenstates of 
the neutral $B^0$ mesons as 
\[
| B_{1,2} \rangle \equiv {1\over \sqrt{1 + | \eta |^2 }}~
\left[ | B^0 \rangle \pm \eta | \overline{B^0} \rangle \right],
\]
with $\eta \equiv \sqrt{(M_{12}^* - i \Gamma_{12}^* )/ 
(M_{12} - i \Gamma_{12} )}$, we can derive the following relation: 
 \begin{equation}
\label{eq:all}
A_{ll} \equiv {N(BB) - N(\bar{B}\bar{B}) \over N(BB) + N(\bar{B}\bar{B})}
= - {| \eta |^4 - 1 \over | \eta |^4 + 1 }
= {{\rm Im} (\Gamma_{12} / M_{12} ) \over 1 + | \Gamma_{12} / M_{12} |^2 / 4 }
\approx {\rm Im} (\Gamma_{12} / M_{12} ).
\end{equation}
Here $M_{12}, \Gamma_{12}$ are the matrix elements of the Hamiltonian in the
$(B^0, \overline{B^0} )$ basis:
%\begin{equation}
%H = \left( \begin{array}{cc}
%           M - i \Gamma /2   &   M_{12} - i \Gama_{12} /2 \\
%           M_{12}^* - i \Gamma_{12}^*   &    M - i \Gamma /2
%           \end{array} \right).  
%\end{equation} 
\[
{1\over 2 m_B}~\langle \overline{B} | H_{\rm full} | B \rangle = 
M_{12} - {i \over 2} \Gamma_{12}.
\]
%One can also show that 
%\[
%A_{ll} \approx {4 {\rm Re} (\epsilon_B) \over 1 + | \epsilon_B |^2 },
%\] 
%which implies that the dilepton charge asymmetry $A_{ll}$ in fact measures 
%the CP violation in the mixing, as the parameter $\epsilon_K$ in the neutral 
%kaon system. 

In the SM, the phases of $M_{12}$ and $\Gamma_{12}$ are approximately equal
and
\[
\Delta M_{\rm SM} \approx 2 | M_{12}^{\rm SM} |, ~~~~
\Delta \Gamma_{\rm SM} \approx 2 | \Gamma_{12}^{\rm SM} |.
\]
the quantity $\Gamma_{12}^{\rm SM}$ is given by \cite{Buras:1997fb}
\begin{eqnarray}
  \Gamma_{12}^{\rm SM} &=&
  (-1) \:
  \frac{G_F^2\:m_b^2\:M_{B_d}\:B_{B_d}\:f_{B_d}^2}{8\pi} \:
  \left[
    v_t^2 + \frac{8}{3}\:v_c\:v_t \left(
      z_c + \frac{1}{4} z_c^2 - \frac{1}{2} z_c^3
    \right) +
  \right. \nonumber\\
  & &
  \left.
    v_c^2
    \left\{
      \sqrt{1 - 4 z_c} \left( 1 - \frac{2}{3} z_c \right) +
      \frac{8}{3} z_c + \frac{2}{3} z_c^2 - \frac{4}{3} z_c^3 - 1
    \right\}
  \right] ,
\end{eqnarray}
where \(v_i \equiv V_{ib}\:V_{id}^* \) and \( z_c \equiv m_c^2 / m_b^2 \).
Varying $f_{B_d}$, $|V_{ub}|$, and $|V_{cb}|$ in the range 
quoted in Table~\ref{tab:input}, and $\gamma$ inside the range given by  
$(54.8 \pm 6.2)^\circ$ \cite{Ciuchini:2000de}, we get the SM value to be
\[
% -1.28 \times 10^{-3} \leq A_{ll}^{\rm SM} \leq -0.48 \times 10^{-3},
-1.54 \times 10^{-3} \leq A_{ll}^{\rm SM} \leq -0.16 \times 10^{-3},
\] 
whereas the current world average is \cite{nir}
\[
A_{ll}^{\rm exp} \approx (0.2 \pm 1.4) \times 10^{-2}.
\]

\begin{table}%[H] add [H] placement to break table across pages
\caption{\label{tab:input}Input values for the parameters.}
\begin{ruledtabular}
\begin{tabular}{c|c}
  $m_{B_d}$ & 5.279 GeV \\
  $m_t$ & 174 GeV \\
  $|V_{cb}|$ & $(40.7 \pm 1.9) \times 10^{-3}$ \\
  $|V_{ub}|$ & $(3.61 \pm 0.46) \times 10^{-3}$ \\
  $f_{B_d}$ & $200 \pm 30$ MeV \\
  $\alpha_s (M_Z)$ & 0.119
\end{tabular}
\end{ruledtabular}
\end{table}

In the presence of SUSY, the phases of $M_{12}$ and $\Gamma_{12}$ may be no
longer the same, and potentially a larger dilepton charge asymmetry may be 
possible.
In particular, $M_{12}$ could be affected strongly by SUSY particles, 
whereas $\Gamma_{12}$ is not, % (since it would be at higher order) :
i.e. $M_{12}^{\rm FULL} = M_{12}^{\rm SM} + M_{12}^{\rm SUSY}$ whereas
$\Gamma_{12}^{\rm FULL} \approx \Gamma_{12}^{\rm SM}$.  In this case,
the dilepton charge asymmetry could be approximated as
\begin{equation}
A_{ll} = {\rm Im} \left( { \Gamma_{12}^{\rm SM} \over
M_{12}^{\rm SM} + M_{12}^{\rm SUSY} } \right).
\end{equation}
The possible ranges of $A_{ll}$ in a class of general SUSY models were 
studied in Ref.~\cite{randall}. 

%%%%%%%%%%%%%%%%%%%%%%%%%%%%%%%%%%%%%%%%%%%%%%%%%%%%%%%%%%%%%%%%%%%%%%%%%%%
\subsection{Effective Hamiltonian for $\Delta B = 1$ processes}
% $B \rightarrow X_d \gamma$}
%%%%%%%%%%%%%%%%%%%%%%%%%%%%%%%%%%%%%%%%%%%%%%%%%%%%%%%%%%%%%%%%%%%%%%%%%%%

The effective Hamiltonian relevant to $\Delta B = 1$ processes involves 
four quark operators and $b\rightarrow d \gamma$ and $b\rightarrow d g$ 
penguin operators. 
Since we are not going to discuss $\Delta B =1$ nonleptonic decays due to
theoretical uncertainties related with factorization, we shall consider
the inclusive radiative decay $B\rightarrow X_d \gamma$ only. 
The relevant effective Hamiltonian for this process is given by \cite{ali} 
\begin{eqnarray}
{\cal H}_{\rm eff} ( b \rightarrow d \gamma (+g)) 
& = & - {4 G_F \over \sqrt{2}}
V_{td}^* V_{tb}~\sum_{i=1,2,7,8}~C_i (\mu_b) O_{ic} (\mu_b)
\nonumber  \\
& + &   {4 G_F \over \sqrt{2}}
V_{ud}^* V_{ub}~\sum_{i=1,2}~C_i (\mu_b) ~
\left[ O_{iu} (\mu_b) - O_{ic} (\mu_b)  \right] 
\end{eqnarray}
with
\begin{eqnarray}
  O_{1c} = \overline{d}_L \gamma^{\mu} c_L~\overline{c}_L \gamma_\mu b_L ,
&& 
  O_{1u} = \overline{d}_L \gamma^{\mu} u_L~\overline{u}_L \gamma_\mu b_L ,
\nonumber 
\\
  O_{2c} = \overline{d}_L \gamma^{\mu} c_L~\overline{c}_L \gamma_\mu b_L ,
&& 
  O_{2u} = \overline{d}_L \gamma^{\mu} u_L~\overline{u}_L \gamma_\mu b_L ,
\nonumber
\\
O_{7\gamma} =  {e \over 16 \pi^2} m_b ~
\overline{d}_L \sigma^{\mu\nu} F_{\mu\nu}  b_R ,
&& 
%\nonumber
%\\
O_{8g}  =  {g_s \over 16 \pi^2} m_b ~
\overline{d}_L \sigma^{\mu\nu} t^a G_{\mu\nu}^a  b_R .
\end{eqnarray}
Here the renormalization scale $\mu_b$ is of the order of $m_b$, and 
we have used the unitarity of the CKM matrix elements 
\[
V_{cd}^* V_{cb} = - ( V_{ud}^* V_{ub} + V_{td}^* V_{tb}),
\]
which should be valid even in the presence of SUSY flavor violations. 

In the SM, all the three up-type quarks contribute to this decay, since
all the relevant CKM factors are of the same order of magnitude.
The strong phases are provided by the imaginary parts of one loop diagrams
at the order $O( \alpha_s )$  by the usual unitarity argument. 
Varying $f_{B_d}$, $|V_{ub}|$, and $|V_{cb}|$ in the range 
quoted in Table~\ref{tab:input}, and $\gamma$ between $(54.8 \pm 6.2)^\circ$
\cite{Ciuchini:2000de}, 
we get the branching ratio for this decay in the SM to be
% $6.0 \times 10^{-6} - 2.6 \times 10^{-5}$ 
$8.9 \times 10^{-6} - 1.1 \times 10^{-5}$.
The direct CP asymmetry in the SM is about
% $-7 \% \sim -35 \%$
$-15 \% - -10 \%$ \cite{ali}. We have updated the previous 
predictions by Ali et al. \cite{ali} using the present values of CKM 
parameters.

The CP averaged branching ratio for $B\rightarrow X_d \gamma$ in the leading
log approximation is given by \cite{ali,kn,keum}
\begin{equation}
{B ( B \rightarrow X_d \gamma ) \over B( B \rightarrow X_c e \nu ) } 
= \left| {V_{td}^* V_{tb} \over V_{cb}} \right|^2 ~{6 \alpha \over \pi f(z) }~
| C_7 ( m_b ) |^2. 
\end{equation} 
where $f(z) = 1 - 8 z + 8 z^3 - z^4 - 12 z^2 \ln z$ is the phase space factor
for the $b\rightarrow c$ semileptonic decays and $\alpha^{-1} = 137.036$.
Neglecting the RG running between the heavy SUSY particles and the top quark 
mass scale, we get the following relations :
\begin{eqnarray}
C_7 ( m_b ) & \approx & - 0.31 + 0.67 ~C_7^{\rm new} ( m_W ) + 
0.09 ~C_8^{\rm new} ( m_W ), 
\nonumber   \\
C_8 ( m_b ) & \approx & - 0.15 + 0.70~ C_8^{\rm new} ( m_W ). 
\end{eqnarray}
The new physics contributions to $C_2$ are negligible so that we use 
$C_2 ( m_b )= C_2^{\rm SM} ( m_b ) \approx 1.11$. 

In general SUSY models considered in the present work, the Wilson 
coefficients $C_{7\gamma}^{\rm new}$ and $C_{8 g}^{\rm new}$ are given by 
\cite{ko2,buras,kane}
\begin{eqnarray}
  C_{7 \gamma}^{\rm SUSY} ( m_W ) & = &  
{8 \pi Q_b \alpha_s \over 3 \sqrt{2} G_F \tilde{m}^2 V_{td}^* V_{tb}}
\left[ ( \delta_{13}^d )_{LL}  M_4 (x)  %\right.
%\nonumber  \\
%& & \left. 
- ( \delta_{13}^d )_{LR} \left( {\tilde{m} \sqrt{x} \over m_b} \right) 
4 B_1 (x) \right],
%- ( \delta_{13}^d )_{LR}^{\rm ind}
%\left( {\tilde{m} \sqrt{x} \over m_b} \right) M_2 (x) \right],
\\
  C_{8 g}^{\rm SUSY} ( m_W ) & = &  {2 \pi \alpha_s \over 
\sqrt{2} G_F \tilde{m}^2 V_{td}^* V_{tb}}
\left[ ( \delta_{13}^d )_{LL} \left( {3\over 2} M_3 (x) - {1\over 6} M_4 (x)
\right) \right.
\nonumber  \\
& & \left.
+ ( \delta_{13}^d )_{LR} \left( {\tilde{m} \sqrt{x} \over m_b} \right) 
~{1\over 6}~\left( 4 B_1 (x) - 9 x^{-1} B_2 (x) \right) \right]  
\nonumber  \\
%& & \left. 
%- ( \delta_{13}^d )_{LR}^{\rm ind}
%\left( {\tilde{m} \sqrt{x} \over m_b} \right)
%\left( {3\over 2} M_1 (x) - {1\over 6} M_2 (x) \right) \right].
\end{eqnarray}
Here we have ignored the RG running between the squark and the gluino mass 
and the $m_W$ scale. 
Note that the $(\delta_{13}^d )_{LR}$ contribution is enhanced by 
$m_{\tilde{g}} / m_b$ compared to the contributions from the SM and 
the $LL$ insertion due to the chirality flip from the internal gluino 
propagator in the loop. 
Explicit expressions for the loop functions $B_i$'s and $M_i$'s can be found 
in Ref.~\cite{ko2,buras,kane}. 

In order to generate a nonvanishing direct CP asymmetry, one needs at least 
two independent amplitudes with different strong (CP-even) and weak (CP-odd) 
phases.  In $B\rightarrow X_d \gamma$, strong phases are provided by
quark and gluon loop diagrams, whereas weak phases are provided  %the Wilson
%coefficients of the effective lagrangian in terms of %$\phi_{1,2,3}$ and
by the KM angles ($\alpha,\beta,\gamma$) and $(\delta_{13}^d )_{AB}$. 
The resulting direct CP asymmetry in $B\rightarrow X_d \gamma$ can be 
written as \cite{ali,kn}
\begin{eqnarray}
A_{\rm CP}^{b\rightarrow d\gamma} ({\rm in} ~\% ) & = & 
{1\over |C_7|^2}~\left[ 10.57 ~{\rm Im} \left( C_2 C_7^* \right) 
- 9.40 ~{\rm Im} \left( ( 1 + \epsilon_d ) C_2 C_7^* \right)   \right.
\nonumber   \\ 
& - & \left. 
9.51 ~{\rm Im} \left( C_8 C_7^* \right) + 0.12 ~{\rm Im} \left(
( 1 + \epsilon_d ) C_2 C_8^* \right) \right],
\label{eq:acp}
\end{eqnarray}
where 
\[
\epsilon_d \equiv {V_{ud}^* V_{ub} \over V_{td}^* V_{tb}}  
\approx { ( \rho - i \eta ) \over  ( 1 - \rho + i \eta )}
\] 
in the Wolfenstein parametrization for the CKM matrix elements.

A remark is in order for the above CP asymmetry in $B\rightarrow X_d \gamma$.
Unlike the $B\rightarrow X_s \gamma$ case for which the $|C_{7\gamma}|$ is 
constrained by the observed  $B\rightarrow X_s \gamma$ branching ratio, 
the  $B\rightarrow X_d \gamma$ decay has not been observed yet, and its 
branching ratio can be vanishingly small even in the presence of new physics.
In that case, $|C_{7\gamma}| \approx 0$ so that the denominator of 
$A_{\rm CP}^{b\rightarrow d\gamma}$ becomes zero and the CP asymmetry blows 
up. This could be partly cured by replacing the denominator $|C_{7\gamma}|^2$
by $K_{\rm NLO} (\delta)$ defined in Ref.~\cite{kn}:
\begin{eqnarray}
K_{\rm NLO} (\delta) ({\rm in} \%) 
& = & 0.11 ~| C_2 |^2 + 68.13 ~| C_7 |^2 + 0.53 ~| C_8 |^2 
%\nonumber  \\
- 16.55 ~{\rm Re}( C_2 C_7^* ) 
\nonumber  \\
& - & 0.01 ~{\rm Re}( C_2 C_8^* ) + 8.85 ~ {\rm Re}( C_7 C_8^* ) 
+ 3.86 ~ {\rm Re}( C_7^{(1)} C_7^* )
\end{eqnarray} 
for the photon energy cutoff factor $\delta = 0.3$. Here $C_7^{(1)}$ is the 
next-to-leading order contribution to $C_{7\gamma} (m_b)$ \cite{kn}:
\begin{equation}
C_{7\gamma}^{(1)} \approx 0.48 - 2.29 ~C_7^{\rm new} ( m_W ) 
- 0.12~ C_8^{\rm new} ( m_W ).
\end{equation}
This prescription will render the denominator of (\ref{eq:acp}) to be finite.
%Also we adopt the upper limit on $B( B \rightarrow X_s g) < 6.8 \%$ as 
%the upper limit on $B( B \rightarrow X_d g)$ in order to put an upper limit 
%on  $|C_8 ( m_b )|^2$:
%\begin{equation}
%B( B \rightarrow X_d g) = \lambda^2 \left[ ( 1 - \rho )^2 + \eta^2 \right] 
%~| C_8 ( m_b ) |^2~B ( B \rightarrow X_c e \nu ). 
%\end{equation} 

%%%%%%%%%%%%%%%%%%%%%%%%%%%%%%%%%%%%%%%%%%%%%%%%%%%%%%%%%%%%%%%%%%%%%%%%%%%
%\section{Physical Observables}
%%%%%%%%%%%%%%%%%%%%%%%%%%%%%%%%%%%%%%%%%%%%%%%%%%%%%%%%%%%%%%%%%%%%%%%%%%%
%\susection{$\Delta m_{B_d}$ and $\sin}

%%%%%%%%%%%%%%%%%%%%%%%%%%%%%%%%%%%%%%%%%%%%%%%%%%%%%%%%%%%%%%%%%%%%%%%%%%%
\section{Numerical Analysis}
%%%%%%%%%%%%%%%%%%%%%%%%%%%%%%%%%%%%%%%%%%%%%%%%%%%%%%%%%%%%%%%%%%%%%%%%%%%

In the numerical analysis, we impose the following quantities as constraints :
\begin{itemize}
\item $\Delta m_{B_d} = (0.472 \pm 0.017) ~{\rm ps}^{-1}$ : This constrains
the modulus of $M_{12} ( B^0 )$ through the following relation
$\Delta m_{B_d} = 2 ~| M_{12} ( B^0 ) |$ \cite{pdg}. 
\item $A_{\rm CP}^{\rm mix} =  ( 0.79 \pm 0.10 )$ :  This constrains
the phase $2 \beta^{'}$ of  $M_{12} ( B^0 )$ by 
$A_{\rm CP}^{\rm mix} = \sin 2 \beta^{'}$, where $2 \beta^{'}$ is the 
argument of $M_{12} ( B^0 )$ \cite{nir}. 
\item $Br ( B \rightarrow X_d \gamma ) < 1 \times 10^{-5}$: 
%Since the current measurement of $B\rightarrow X_s \gamma$ agrees well with 
%the SM prediction, there is very little room for new physics contributions 
%to this decay. 
%By the same token, the decay $B \rightarrow X_d \gamma$ is also 
%strongly constrained. 
At present, there are limits only on the exclusive decays: 
$B( B\rightarrow \rho \gamma) < 2.3 \times 10^{-6}$. We assume 
a modest upper bound on the branching ratio for the inclusive radiative decay 
as $Br ( B \rightarrow X_d \gamma ) \lesssim 1 \times 10^{-5}$.
This is also well below the experimental uncertainty in the    
$B\rightarrow X_s \gamma$ branching ratio.
This puts a strong constraint on both $LL$ and $LR$ insertions as we 
shall see. Especially the $LR$ insertion is more strongly constrained, since 
its contribution is enhanced by $m_{\tilde{g}} / m_b$ due to the chirality 
flip from the gluino in the loop compared to other contributions including 
the SM one. This is a new ingredient compared to the work in 
Ref.~\cite{masiero2002}.
\item $A_{ll}^{\rm exp} = (0.2 \pm 1.4) \%$: This is related to the 
$B^0 - \overline{B^0}$ mixing through the relation (\ref{eq:all}). Although 
we do not use this constraint to restrict the allowed parameter space, we 
indicate the parameter space where the resulting $A_{ll}$ falls out of the 
$1\sigma$ range. It turns out that both $LL$ and $LR$ mixing scenarios
are already strongly constrained by the $B\rightarrow X_d \gamma$ branching 
ratio  rather than by $A_{ll}$, as can be seen in Figs.~2 (a) and (b).
\end{itemize}
We impose these constraints at 68 \% C.L. ($1 \sigma$) as we vary the KM
angle $\gamma$ between 0 and $2 \pi$. In all cases, we set the common squark 
mass $\tilde{m} = 500$ GeV and $x = 1$ ($m_{\tilde{g}} = \tilde{m}$).
Finally for the mass insertion parameters $(\delta_{13}^d )_{AB}$, we 
consider two cases.
In the first case (the $LL$ case), only $(\delta_{13}^d )_{LL}$
is nonvanishing among the mass insertion parameters,
and in the second (the $LR$ case), only $(\delta_{13}^d )_{LR}$.
It would be straightforward to consider other possibilities such as 
$( \delta_{13}^d )_{LL} = ( \delta_{13}^d )_{RR}$ etc. in a similar way. 

% four cases as in Ref.~X : (i) $(\delta_{13}^d )_{LL}$ only, 
% (ii) $(\delta_{13}^d )_{LL} =(\delta_{13}^d )_{RR}$, 
% (iii) $(\delta_{13}^d )_{LR}$ only, 
% (iv) $(\delta_{13}^d )_{LR} =(\delta_{13}^d )_{RL}$. 
% [IS IT RALLY NECESSARY ? WE CAN DO ONLY (i) and (iii), SINCE IT IS 
% A SORT OF QUALITATIVE ANALYSIS ANYWAY.]

\begin{figure}
\centering
\subfigure[$LL$ mixing only]{\includegraphics[width=8cm]{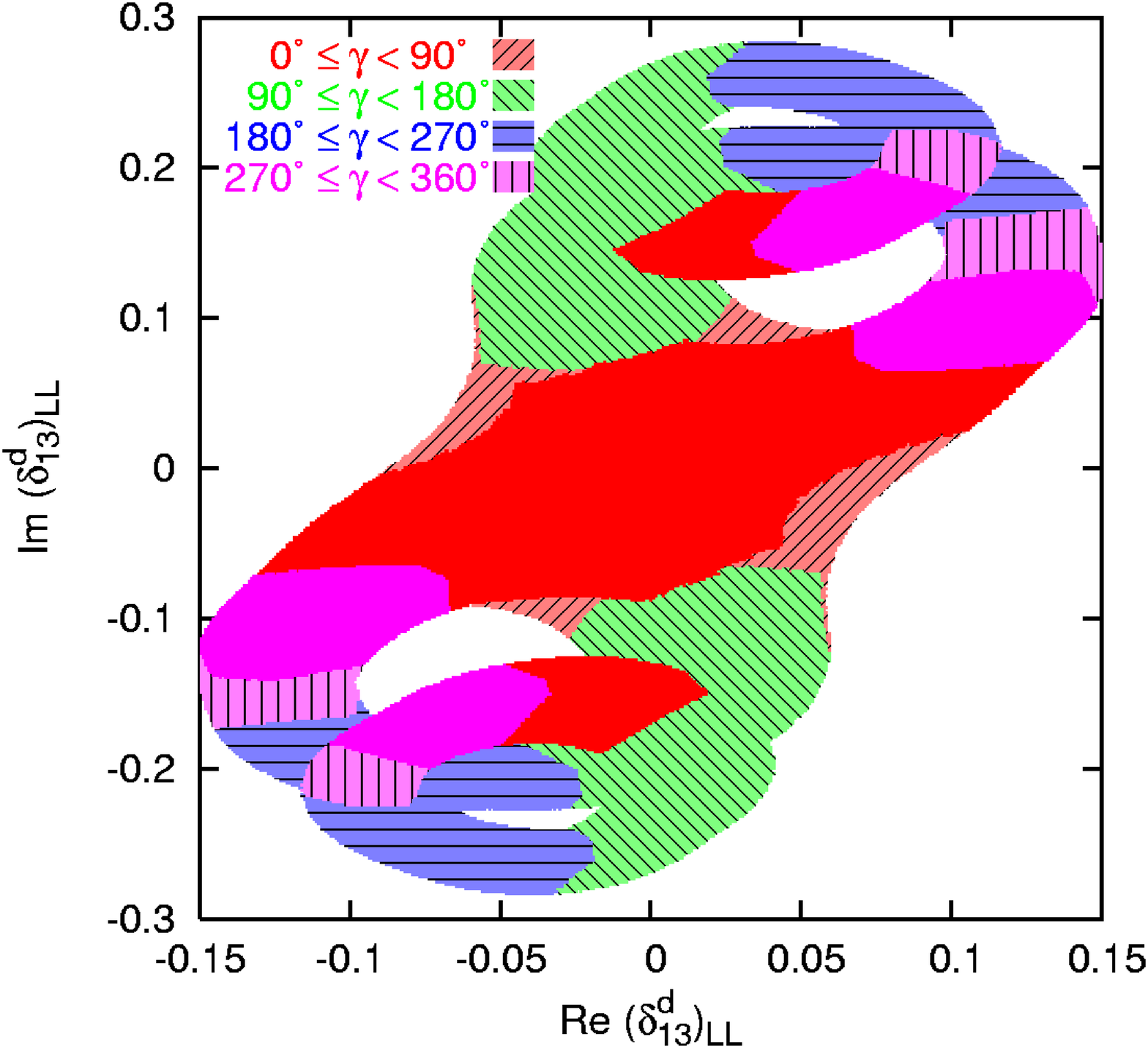}}
%\subfigure[$LL$ mixing only]{\includegraphics[width=8cm]
%{all-LL-0TeV-cropped.eps}}
\subfigure[$LR$ mixing only]{\includegraphics[width=8cm]{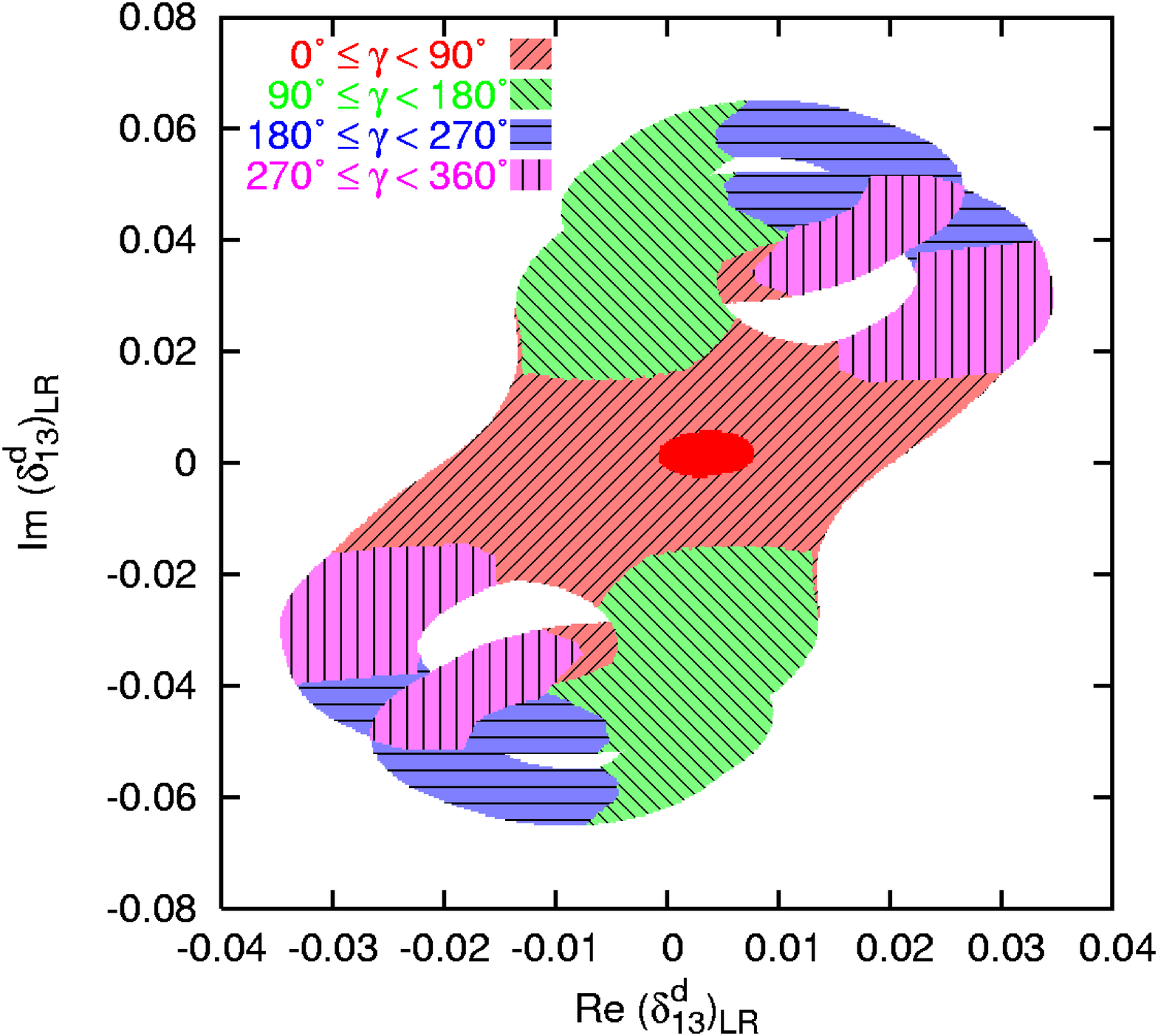}}
\caption{
The allowed ranges in (a) the $LL$ and (b) the $LR$ insertion cases for 
the parameters 
$( {\rm Re}(\delta_{13}^d )_{AB}, {\rm Im} (\delta_{13}^d )_{AB})$ for 
different values of the KM angle $\gamma$ with different color codes:
dark (red) for $0^{\circ} \leq \gamma \leq 90^{\circ}$, light gray (green)
for  $90^{\circ} \leq \gamma \leq 180^{\circ}$, very dark (blue) for
$180^{\circ} \leq \gamma \leq 270^{\circ}$ and gray (magenta) for 
$270^{\circ} \leq \gamma \leq 360^{\circ}$.
% In (a), the region leading to a too large $A_{ll}$ is covered by the dots.
% In (b), 
The region leading to a too large branching ratio for 
$B_d \rightarrow X_d \gamma$ is colored lightly and covered by parallel lines.
}
\label{fig:d13}
\end{figure}

The parameter space searching is done in the following way.
%\begin{itemize}
%\item 
We vary $\gamma$ from 0 to $2\pi$, and $(\delta^d_{13})_{AB}$
  inside the bound depicted in Ref.~\cite{masiero2002}.
%\item 
For a given set of values of $\gamma$ and $(\delta^d_{13})_{AB}$,
  we search for those $f_{B_d}$ and 
  $\sqrt{\rho^2 + \eta^2} \equiv |V_{ub}|/ \lambda V_{cb}$ (with $\lambda =
  | V_{us} |$) 
  that satisfy 1-$\sigma$ constraints on $\Delta M_B$ and $\sin 2 \beta'$.
  The search region is the 1-$\sigma$ range in Table~\ref{tab:input}.
%\item 
If no such pair exists, the $(\gamma, (\delta^d_{13})_{AB})$
  point is excluded from the plots.
  Points that are not excluded are plotted in Fig.~\ref{fig:d13}.
%\item 
Using these $\gamma$, $(\delta^d_{13})_{AB}$,
  $f_{B_d}$, and $\sqrt{\rho^2 + \eta^2}$ found above,
  we plot $Br ( B \rightarrow X_d \gamma )$
  and $A_{\rm CP}^{b\rightarrow d\gamma}$.
%  Those parameters that lead to %$B ( B \rightarrow X_d \, g ) > 6.8 \%$
%  are marked with light gray (magenta),
%  $B(B\rightarrow X_d \, g) < 6.8 \%$ but 
%  $B ( B \rightarrow X_d \gamma ) > 0.5 \times 10^{-4}$
%  is marked with gray (magenta), 
%  and those for $B ( B \rightarrow X_d \gamma ) < 0.5 \times 10^{-4}$
%  with black (on a color display).
%\end{itemize}
Uncertainties in $B_{1-3} (m_b)$, which are actually used
in our analysis, are only several per cent, while
that in $f_{B_d}$ is $15 \%$.
Moreover the matrix elements (\ref{eq:db2me})
are proportional to $f_{B_d}^2 B_i (m_b)$, so
we do not take into account the uncertainties in the bag parameters.

\begin{figure}
\centering
\subfigure[$LL$ mixing only]{\includegraphics[width=8cm]%
{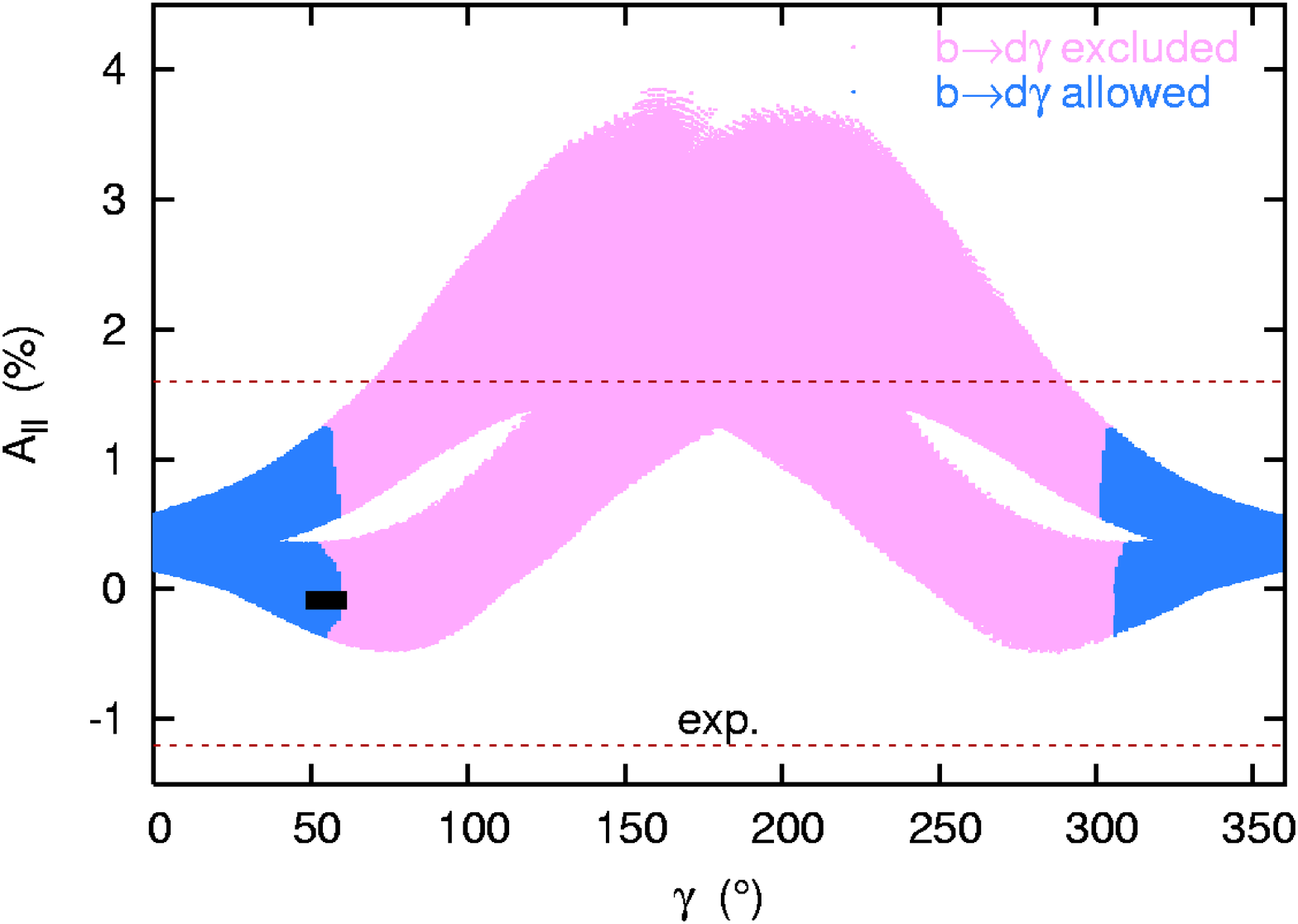}}
%\subfigure[$LR$ mixing only]{\includegraphics[width=8cm]{LR2d-ex2.eps}}
\subfigure[$LR$ mixing only]{\includegraphics[width=8cm]
{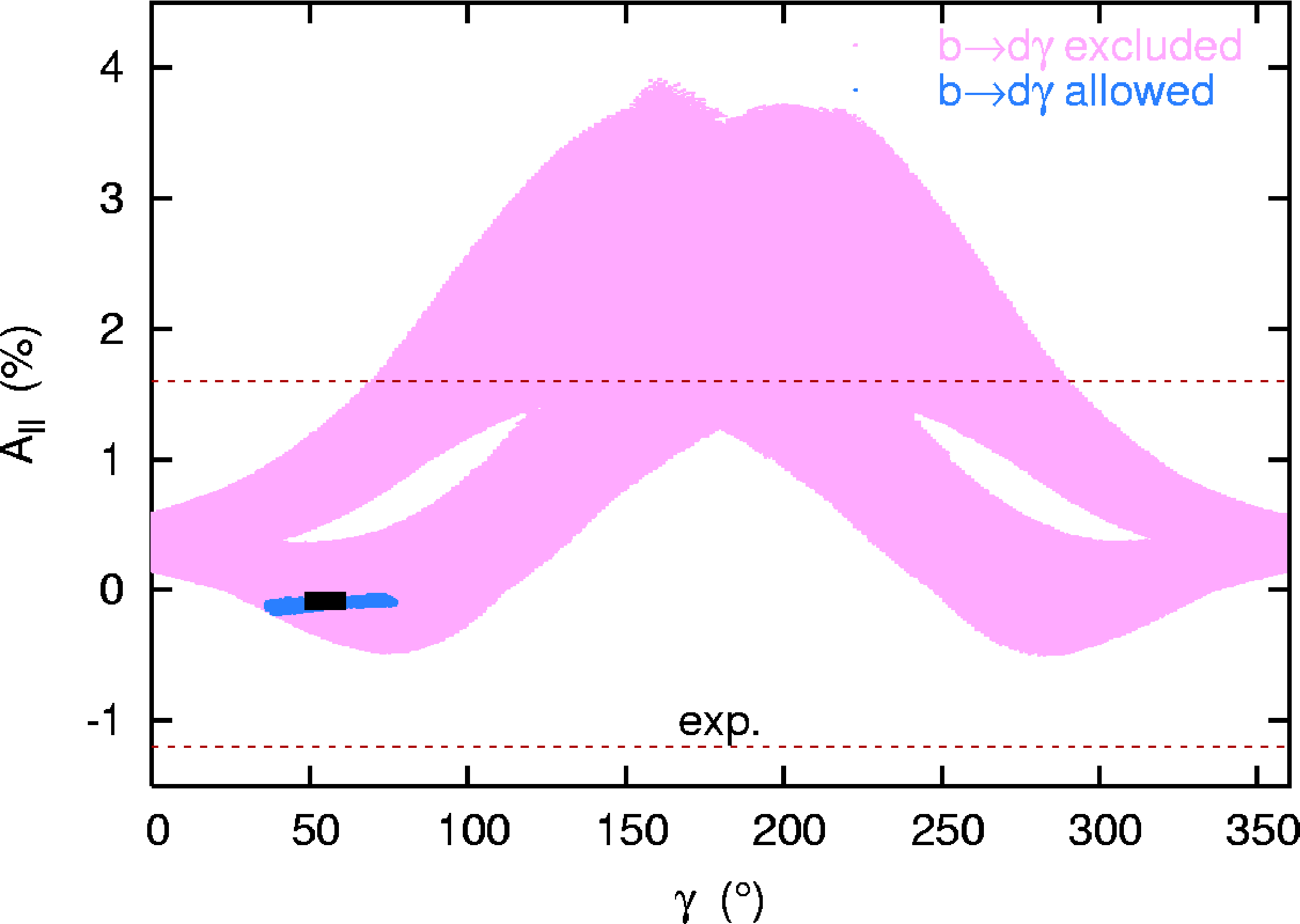}}
%\subfigure[$\tan \beta = 3$]{%
%\includegraphics[height=7cm]{xxx.eps}}\hspace{5mm}
%\subfigure[$\tan \beta = 30$]{%
%\includegraphics[height=7cm]{xxx.eps}}
\caption{
The possible ranges of the dilepton charge asymmetry
in (a) the $LL$ and (b) the $LR$ cases
as functions of the KM angles $\gamma$. 
  The black rectangle around $\gamma \simeq 55^\circ$ is
  the SM prediction.
Those parameters which lead to
$B ( B \rightarrow X_d \gamma ) > 1 \times 10^{-5}$ are denoted by the
the gray (magenta) region, and those for 
$B ( B \rightarrow X_d \gamma ) < 1 \times 10^{-5}$ by the 
dark (blue) region. The $1 \sigma$ range for the world average of 
$A_{ll}^{\rm exp} =  (0.2 \pm 1.4) \%$ is shown to lie between 
the short dashed lines.
}
\label{fig:all}
\end{figure}

In Figs.~\ref{fig:d13} (a) and (b), we show the allowed parameter space in 
the $( {\rm Re}(\delta_{13}^d )_{AB}, {\rm Im} (\delta_{13}^d )_{AB})$ plane 
[(a)  $LL$ insertion and (b) $LR$ insertion, respectively]   
for different values of the KM angle $\gamma$ with different color codes:
dark (red) for $0^{\circ} \leq \gamma \leq 90^{\circ}$, light gray (green)
for  $90^{\circ} \leq \gamma \leq 180^{\circ}$, very dark (blue) for
$180^{\circ} \leq \gamma \leq 270^{\circ}$ and gray (magenta) for 
$270^{\circ} \leq \gamma \leq 360^{\circ}$. The region leading 
to a too large branching ratio for $B_d \rightarrow X_d \gamma$ is covered 
by slanted lines. And the region where $A_{ll}$ falls out of the data within
$1\sigma$ range is already excluded by the $B\rightarrow X_d \gamma$ 
branching ratio constraint. For both the $LL$ and $LR$ mixing cases,
our results are the same as those in Ref.~\cite{masiero2002},
if we impose only the $\Delta m_{B_d}$ and $\sin 2 \beta$ constraints. 
By adding a constraint from $B\rightarrow X_d \gamma$ (and $A_{ll}$),
the allowed parameter space is further reduced, and the effect is even larger 
for the $LR$ mixing case. For the $LL$ mixing [ Fig.~1 (a) ], 
the $B\rightarrow X_d \gamma$ does play some role,  %On the other hand, 
and the $A_{ll}$ gives a moderate constraint. The KM angle $\gamma$ should be
in the range between  $\sim - 60^\circ$ and $\sim + 60^{\circ}$, and $A_{ll}$
can have the opposite sign compared to the SM prediction, even if the KM 
angle  is the same as its SM value $\gamma \simeq 55^{\circ}$.  
For the $LR$ mixing [ Fig.~1 (b) ], 
the $B(B_d \rightarrow X_d \gamma)$ puts an even stronger constraint on 
the $LR$ insertion, whereas the $A_{ll}$ does not play any role. 
In particular, the KM  angle $\gamma$ can not be too much different from 
the SM value in the $LR$ mixing case, once the 
$B(B_d \rightarrow X_d \gamma)$  constraint is included. 
Only $30^{\circ} \lesssim \gamma \lesssim 80^{\circ}$ is compatible with all 
the data from the $B$ system, even if we do not consider the $\epsilon_K$ 
constraint. The resulting parameter space is significantly reduced compared 
to the result obtained in Ref.~\cite{masiero2002}. 
The limit on the $LR$ insertion parameter will become even stronger as the
experimental limit on $B_d \rightarrow X_d \gamma$ will be improved in the 
future. %On the other hand, for the $LL$ mixing case one can have 
%$( \delta_{13}^d )_{LL} \sim O(0.1)$ for any values of the KM angle $\gamma$
%between $\sim - 60^\circ$ and $\sim + 60^{\circ}$. 

In Fig.~\ref{fig:all}, we show the predictions for $A_{ll}$ as a function of 
the KM angle $\gamma$: (a) $LL$ insertion and (b) $LR$ insertion only. For 
the $LL$ insertion case [Fig.~2 (a)], one can expect a large deviation from 
the SM prediction for $A_{ll}$ for a wide range of the KM angle $\gamma$
($\sim - 60^{\circ} \lesssim \gamma \lesssim + 60^\circ$),
even after we impose the $B\rightarrow X_d \gamma$ branching ratio which is 
more constraining than the $A_{ll}$ (the short dashed lines  indicate the 
$1\sigma$ range for $A_{ll}^{\rm exp}$). 
Also even if the KM angle $\gamma$ is close to the SM value 
($\gamma \approx 55^{\circ}$), the dilepton charge asymmetry $A_{ll}$ can be 
different from the SM prediction by a significant amount due to the SUSY 
contributions from $( \delta_{13}^3 )_{LL}$. On the other hand, for the 
$LR$ insertion case [Fig.~2 (b)], the $B\rightarrow X_d \gamma$ constraint 
rules out essentially almost all the parameter space region, and the resulting 
$A_{ll}$ is essentially the same as for the SM case.

\begin{figure}
\centering
\subfigure[B ($B\rightarrow X_d \gamma)$]{\includegraphics[width=8cm]
{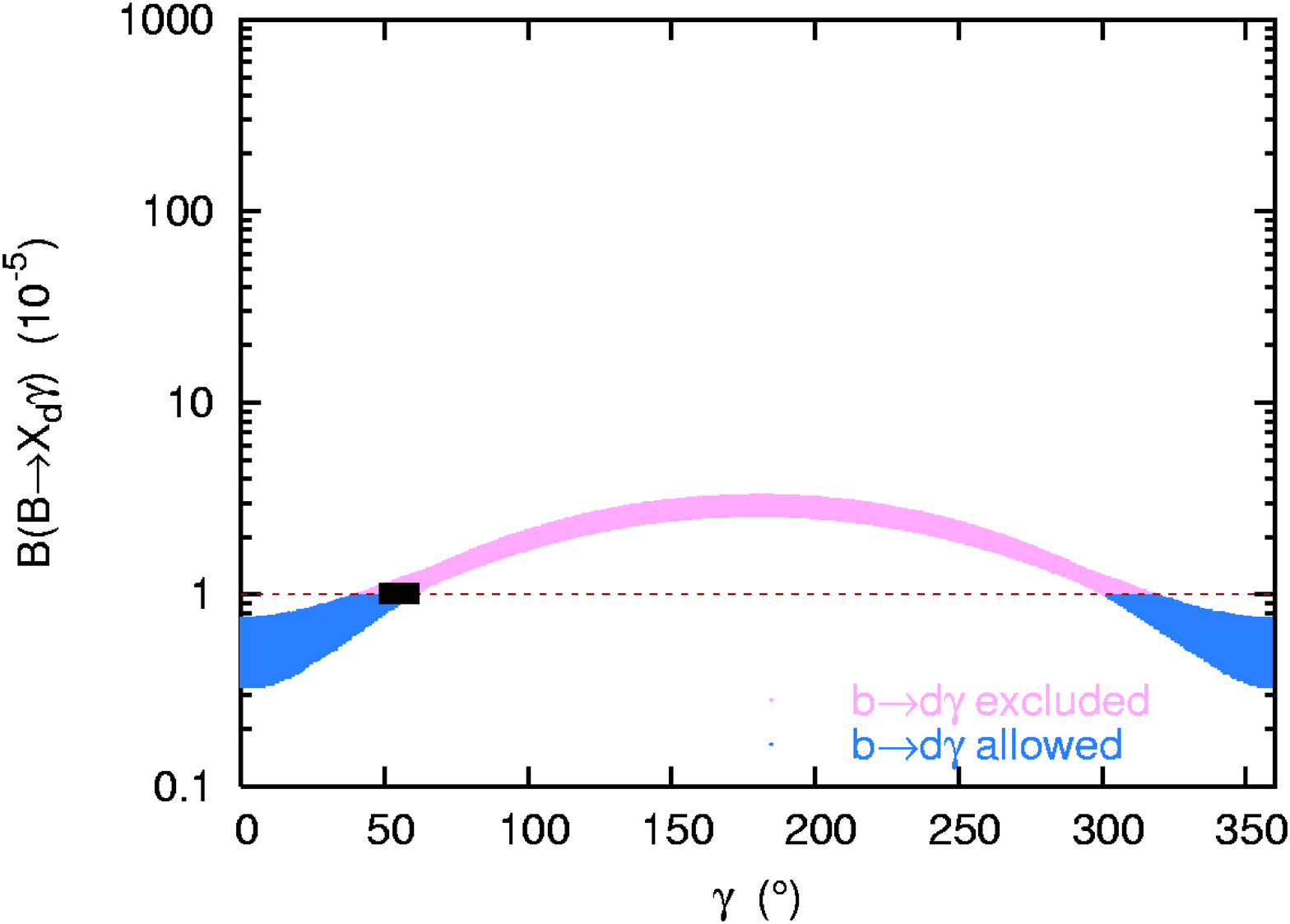}}
%\subfigure[$LR$ mixing only]{\includegraphics[width=8cm]
%{br-LR-0TeV-cropped.eps}}
\subfigure[$A_{\rm CP}^{b\rightarrow d\gamma}$]{\includegraphics[width=8cm]
{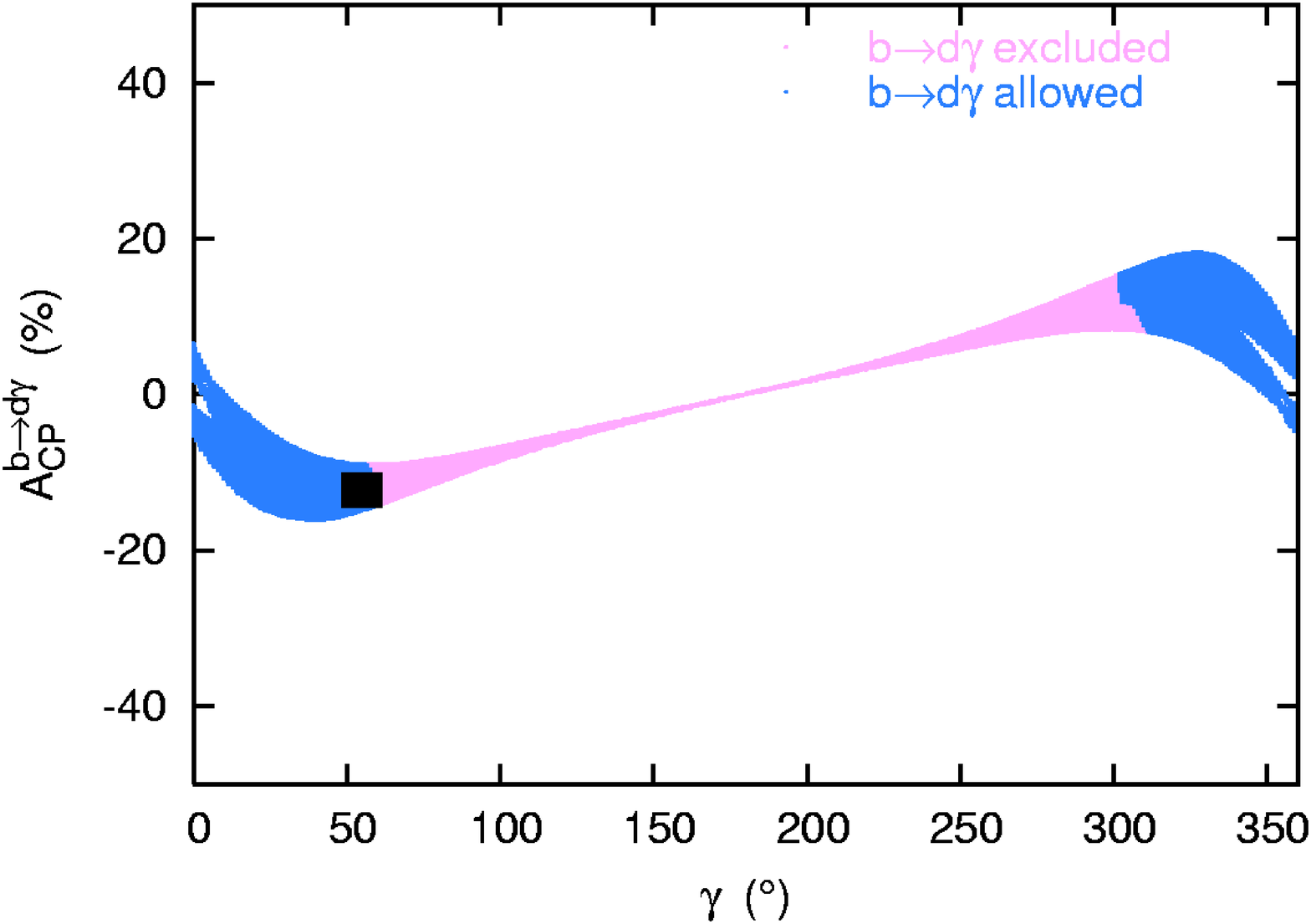}}
%\subfigure[$\tan \beta = 3$]{%
%\includegraphics[height=7cm]{xxx.eps}}\hspace{5mm}
%\subfigure[$\tan \beta = 30$]{%
%\includegraphics[height=7cm]{xxx.eps}}
\caption{
The possible ranges of (a) $B( B_d \rightarrow X_d \gamma )$ and 
(b) %the direct CP asymmetry 
$A_{\rm CP}^{b\rightarrow d\gamma}$ %therein
as functions of the KM angle $\gamma$  in the $LL$ insertion case. 
The black rectangle around $\gamma \simeq 55^\circ$ is
the SM prediction. 
% In the $LL$ mixing case, the 
% $B( B_d \rightarrow X_d \gamma )$ is irrelevant. 
Those parameters which lead to
$B ( B \rightarrow X_d \gamma ) > 1 \times 10^{-5}$ are represented by
the gray (magenta) region, and those for 
$B ( B \rightarrow X_d \gamma ) < 1 \times 10^{-5}$ by the 
dark (blue) region.
}
\label{fig:ll}
\end{figure}

\begin{figure}
\centering
\subfigure[B ($B\rightarrow X_d \gamma)$]{\includegraphics[width=8cm]
{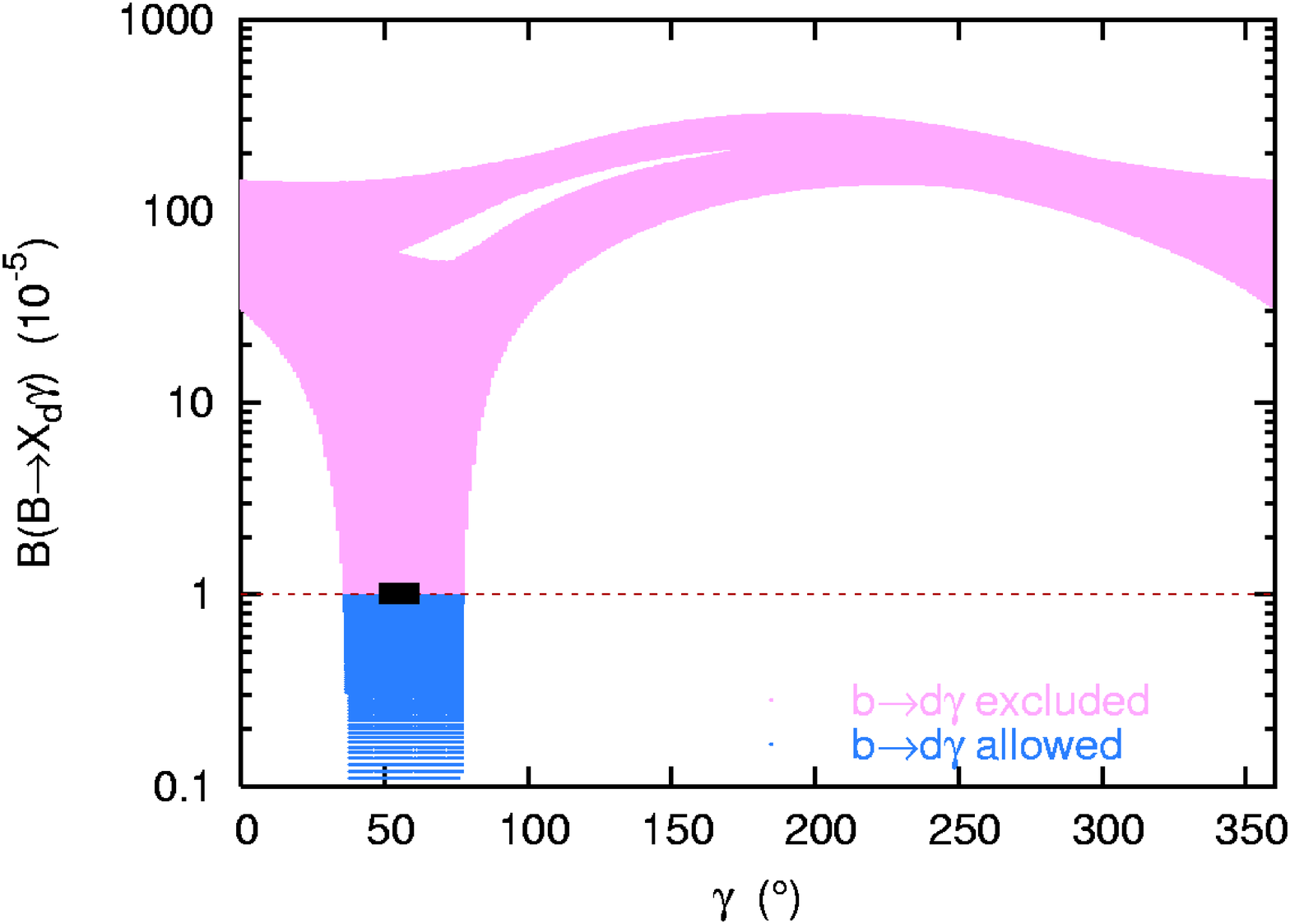}}
\subfigure[$A_{\rm CP}^{b\rightarrow d\gamma}$]{\includegraphics[width=8cm]
{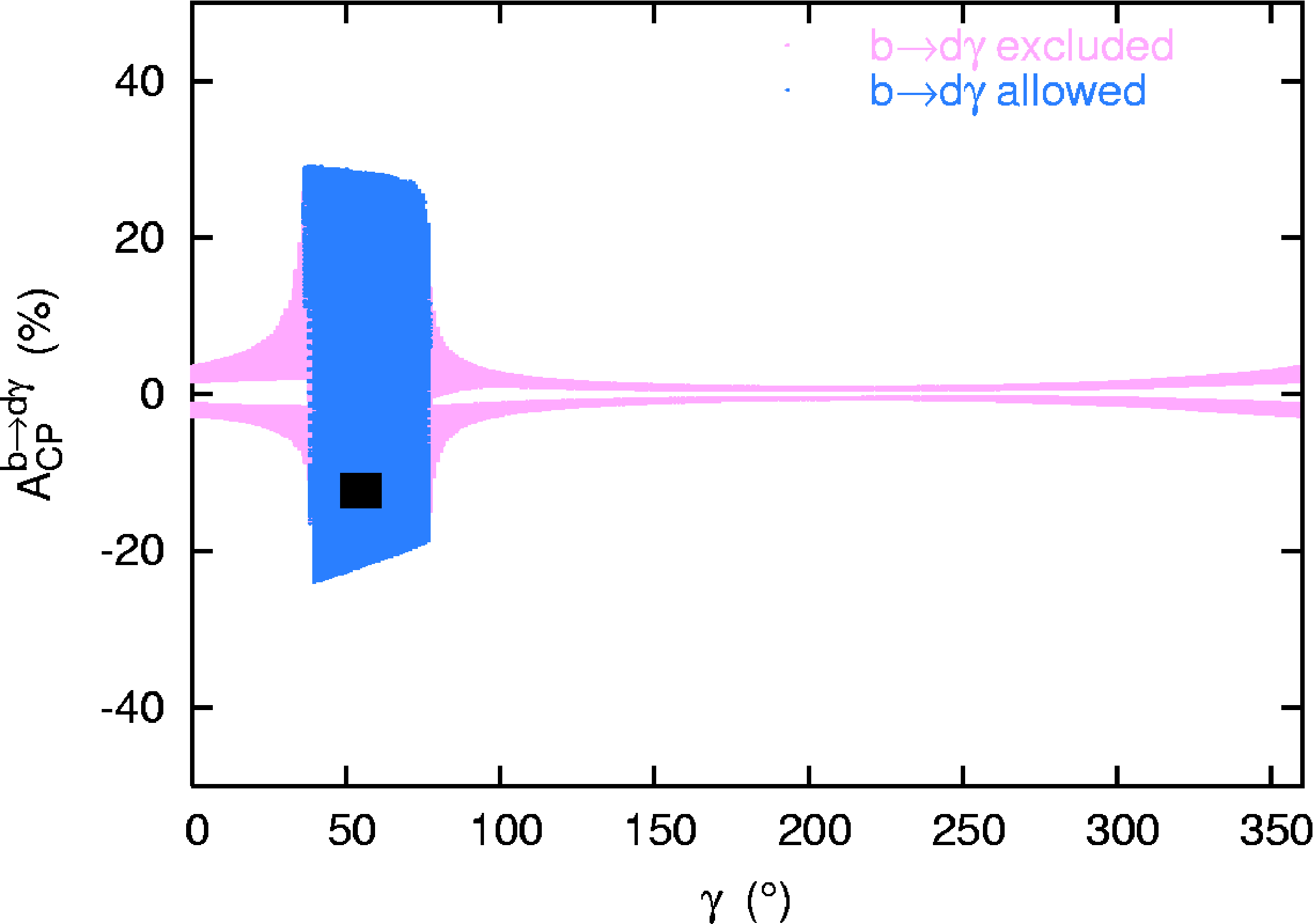}}
%\subfigure[$\tan \beta = 3$]{%
%\includegraphics[height=7cm]{xxx.eps}}\hspace{5mm}
%\subfigure[$\tan \beta = 30$]{%
%\includegraphics[height=7cm]{xxx.eps}}
\caption{
The possible ranges of (a) 
$B (B_d \rightarrow X_d \gamma )$ and 
(b) %the direct CP asymmetry 
$A_{\rm CP}^{b\rightarrow d\gamma}$ %therein  
as functions of the KM angle $\gamma$  in the $LR$ insertion case . 
  The black rectangle around $\gamma \simeq 55^\circ$ is
  the SM prediction.
Those parameters which lead to
$B ( B \rightarrow X_d \gamma ) > 1 \times 10^{-5}$ are represented by
the gray (magenta) region, and those for 
$B ( B \rightarrow X_d \gamma ) < 1 \times 10^{-5}$ by the 
dark (blue) region.
}
\label{fig:lr}
\end{figure}

In Figs.~\ref{fig:ll} (a) and (b),
we show the branching ratio of $B_d \rightarrow X_d \gamma$ and the direct 
CP asymmetry therein, respectively, as functions of the KM angles $\gamma$ 
for the $LL$ insertion only. The SM predictions 
\[
B (B_d \rightarrow X_d \gamma) = (0.9 - 1.1 ) \times 10^{-5},
~~~A_{\rm CP}^{b\rightarrow d \gamma} = (-15 \sim -10) \%
\] 
are indicated by the black boxes. 
In this case, the KM angle $\gamma$ is constrained in the range 
$\sim - 60^\circ$ and $\sim + 60^{\circ}$.
%,  so that the KM angle $\gamma$ can take any values from 0 to $2 \pi$. 
%The resulting $B (B_d \rightarrow X_d \gamma)$ can vary between
%$\sim 0.3\times 10^{-5}$ to $\sim 3.3 \times 10^{-5}$ depending on $\gamma$.
The direct CP asymmetry is predicted to be between $\sim - 15\%$ and 
$\sim +20\%$.
%For $\gamma = 180^{\circ}$, the branching ratio becomes maximal whereas the 
%direct CP asymmetry vanishes. 
In the $LL$ mixing case, the SM gives the dominant contribution to  
$B_d \rightarrow X_d \gamma$, but the KM angle can be different from the 
SM case, because SUSY contributions to the $B^0 - \overline{B^0}$ mixing can 
be significant and the preferred value of $\gamma$ can change from the SM 
KM fitting. This is the same in the rare kaon decays and the results 
obtained in Ref.~\cite{ko2} apply without modifications.
If the KM angle $\gamma$ is substantially different from the SM value (say,
$\gamma = 0$), we could anticipate large deviations in the 
$B_d \rightarrow X_d \gamma$ branching ratio and the direct CP violation
thereof. 

In Figs.~\ref{fig:lr} (a) and (b),
we show the branching ratio of $B_d \rightarrow X_d \gamma$ and the direct 
CP asymmetry therein, respectively, as functions of the KM angles $\gamma$ 
for the $LR$ insertion only. As  before, the black boxes 
represent the SM predictions for $B (B_d \rightarrow X_d \gamma)$ and 
the direct CP asymmetry therein. In the $LR$ insertion case, there could 
be substantial deviations in both the branching ratio and the CP asymmetry 
from the SM predictions, even if the $\Delta m_B$ and $\sin 2 \beta$ is 
the same as the SM predictions as well as the data. For the $LL$ insertion, 
such a large deviation is possible, since the KM angle $\gamma$ can be 
substantially different from the SM value. On the other hand, for the $LR$ 
mixing, the large deviation comes from the complex $( \delta_{13}^d )_{LR}$ 
even if the KM angle is set to the same value as in the SM. 
The size of $( \delta_{13}^d )_{LR}$ is too small to affect the 
$B^0 - \overline{B^0}$ mixing, but is still large enough too affect 
$B\rightarrow X_d \gamma$. Our model independent study indicates that
the current data on the $\Delta m_B$, $\sin2\beta$ and $A_{ll}$ do still 
allow a possibility for large deviations in $B\rightarrow X_d \gamma$,
both in the branching ratio and the direct CP asymmetry thereof. 
The latter variables are indispensable to test completely the KM paradigm for
CP violation and get ideas on possible new physics with new flavor/CP 
violation in $b\rightarrow d$ transition.

%%%%%%%%%%%%%%%%%%%%%%%%%%%%%%%%%%%%%%%%%%%%%%%%%%%%%%%%%%%%%%%%%%%%%%%%%%%%%
\section{Conclusions}
%%%%%%%%%%%%%%%%%%%%%%%%%%%%%%%%%%%%%%%%%%%%%%%%%%%%%%%%%%%%%%%%%%%%%%%%%%%%%

In this work, we considered the gluino-mediated SUSY contributions to 
$B^0 - \overline{B^0}$ mixing, $B\rightarrow J/\psi K_s$ and 
$B\rightarrow X_d \gamma$ in the mass insertion approximation.
We find that the $(LL)$ mixing parameter can be as large as 
$| (\delta_{13}^d)_{LL} | \lesssim 2 \times 10^{-1}$, but the 
$(LR)$ mixing is strongly constrained by the $B\rightarrow X_d \gamma$ 
branching ratio: $| (\delta_{13}^d)_{LR} | \lesssim 10^{-2}$.
The implications for the direct CP asymmetry in $B\rightarrow X_d \gamma$ 
are also discussed, where substantial deviations from the SM predictions are 
possible  both in the $LL$ and $LR$ insertion cases for different reasons. 
For the $LL$ insertion case, the SUSY contribution to 
$B\rightarrow X_d \gamma$ is not so significant, but is still constrained 
by the current upper limit on $B\rightarrow X_d \gamma$. (If the upper limit
were $B(B\rightarrow X_d \gamma ) < 5 \times 10^{-5}$, then the allowed 
region for the KM angle $\gamma$ is the whole range from $0$ to $ 2 \pi$,
as can be seen from Fig.~3 (a). In this case, the $A_{ll}$ will provide a 
more important constraint for the $LL$ insertion.) Also the global 
KM fitting can change because SUSY can affect $B^0 - \overline{B^0}$ 
mixing by a significant manner. By the same reason, there is still ample room
for large deviations in the $A_{ll}$ for the $LL$ insertion case. 
On the other hand, for the $LR$ insertion case, the SUSY
contribution to $B\rightarrow X_d \gamma$ is enhanced by the factor 
$m_{\tilde{g}} / m_b$ and the size of $( \delta_{13}^d )_{LR}$ is strongly 
constrained. The resulting effect is that the KM angle cannot differ too much
from the SM case. Still large deviations in the branching ratio for 
$B\rightarrow X_d \gamma$ and direct CP violation thereof is possible due to
large SUSY loop effects on $B\rightarrow X_d \gamma$.
Thus it turns out that all the observables, $A_{ll}$, the branching ratio 
of $B\rightarrow X_d \gamma$ and the direct CP violation thereof are very 
important, since they could provide informations on new flavor
and CP violation from $(\delta_{13}^d )_{LL,LR}$ (or any other new physics 
scenarios with new flavor/CP violations).
Also they are indispensable in order that we can ultimately test 
the KM paradigm for CP vioaltion in the SM. 

\vspace{.2cm}
\noindent 
{\bf\it Note Added}

While this work was being finished, we recieved a preprint \cite{ali02}, 
in which similar processes (the exclusive  $B\rightarrow \rho \gamma$ 
and various asymmetries thereof, and $A_{ll}$) in certain class of SUSY 
models are considered.

\acknowledgements
This work is supported in part by BK21 Haeksim program of the Ministry of 
Education (MOE), by the Korea Science and Engineering Foundation (KOSEF) 
through Center for High Energy Physics (CHEP) at Kyungpook National 
University, and by DFG-KOSEF Collaboration program (2000) under the contract 
20005-111-02-2 (KOSEF) and 446 KOR-113/137/0-1 (DFG).


\begin{thebibliography}{99}

\bibitem{exp:sin2beta}
B.~Aubert {\it et al.}  [BABAR Collaboration],
%``Improved measurement of the CP-violating asymmetry amplitude  sin(2beta),''
arXiv:hep-ex/0203007;
%%CITATION = HEP-EX 0203007;%%
T.~Higuchi  [Belle Collaboration],
%``Improved measurement of CP asymmetry in the neutral B meson system,''
arXiv:hep-ex/0205020.
%%CITATION = HEP-EX 0205020;%%

\bibitem{masiero2002} D.~Becirevic {\it et al.},
%``B/d anti-B/d mixing and the B/d $\to$ J/psi K(S) asymmetry in general  
%SUSY models,''
arXiv:hep-ph/0112303.
%%CITATION = HEP-PH 0112303;%%

\bibitem{pdg}
D.~E.~Groom {\it et al.}  [Particle Data Group Collaboration],
%``Review Of Particle Physics,''
Eur.\ Phys.\ J.\ C {\bf 15}, 1 (2000).
%%CITATION = EPHJA,C15,1;%%

\bibitem{nir} 
S.~Laplace, Z.~Ligeti, Y.~Nir and G.~Perez,
%``Implications of the CP asymmetry in semileptonic B decay,''
Phys.\ Rev.\ D {\bf 65}, 094040 (2002)
[arXiv:hep-ph/0202010].
%%CITATION = HEP-PH 0202010;%%

\bibitem{kkl}
A.~Ali, G.~F.~Giudice and T.~Mannel,
%``Towards a model independent analysis of rare B decays,''
Z.\ Phys.\ C {\bf 67}, 417 (1995)
[arXiv:hep-ph/9408213];
%%CITATION = HEP-PH 9408213;%%
Y.~G.~Kim, P.~Ko and J.~S.~Lee,
%``Possible new physics signals in b $\to$ s gamma and b $\to$ s l+ l-,''
Nucl.\ Phys.\ B {\bf 544}, 64 (1999)
[arXiv:hep-ph/9810336];
%%CITATION = HEP-PH 9810336;%%
A.~Ali, E.~Lunghi, C.~Greub and G.~Hiller,
%``Improved model-independent analysis of semileptonic and radiative 
% rare  B decays,''
arXiv:hep-ph/0112300.
%%CITATION = HEP-PH 0112300;%%

\bibitem{antichi}
F.~Gabbiani and A.~Masiero,
%``Fcnc In Generalized Supersymmetric Theories,''
Nucl.\ Phys.\ B {\bf 322} (1989) 235 ;
%%CITATION = NUPHA,B322,235;%%
%\cite{Hagelin:1994tc}
%\bibitem{Hagelin:1994tc}
J.~S.~Hagelin, S.~Kelley and T.~Tanaka,
%``Supersymmetric flavor changing neutral currents: Exact amplitudes 
%and phenomenological analysis,''
Nucl.\ Phys.\ B {\bf 415} (1994) 293 ;
%%CITATION = NUPHA,B415,293;%%
E.~Gabrielli, A.~Masiero and L.~Silvestrini,
Phys.\ Lett.\ B {\bf 374} (1996) 80
[arXiv:hep-ph/9509379];
%%CITATION = HEP-PH 9509379;%%
%\cite{Gabbiani:1996hi}
%\bibitem{Gabbiani:1996hi}
F.~Gabbiani, E.~Gabrielli, A.~Masiero and L.~Silvestrini,
%``A complete analysis of FCNC and CP constraints in general SUSY 
%extensions of the standard model,''
Nucl.\ Phys.\ B {\bf 477} (1996) 321
[arXiv:hep-ph/9604387];
%%CITATION = HEP-PH 9604387;%%

\bibitem{ali1}
A.~Ali and E.~Lunghi,
%``Extended minimal flavor violating MSSM and implications for B physics,''
Eur.\ Phys.\ J.\ C {\bf 21}, 683 (2001)
[arXiv:hep-ph/0105200];
%%CITATION = HEP-PH 0105200;%%
A.~Ali, 
Contribution to International Conference on Flavor Physics (ICFP 2001), 
Zhang-Jia-Jie City, Hunan, China, 31 May - 6 Jun 2001.
%``Signatures of supersymmetry in B decays: A theoretical perspective,''
arXiv:hep-ph/0201120.
%%CITATION = HEP-PH 0201120;%%

\bibitem{kane} 
L.~Everett, G.~L.~Kane, S.~Rigolin, L.~T.~Wang and T.~T.~Wang,
%``Alternative approach to b $\to$ s gamma in the uMSSM,''
JHEP {\bf 0201}, 022 (2002)
[arXiv:hep-ph/0112126].
%%CITATION = HEP-PH 0112126;%%

\bibitem{mfv} 
S.~Baek and P.~Ko,
%``Probing SUSY-induced CP violations at B factories,''
Phys.\ Rev.\ Lett.\  {\bf 83}, 488 (1999)
[arXiv:hep-ph/9812229];
%%CITATION = HEP-PH 9812229;%%
A.~Ali and D.~London,
%``Profiles of the unitarity triangle and CP-violating phases in the  standard model and supersymmetric theories,''
Eur.\ Phys.\ J.\ C {\bf 9}, 687 (1999)
[arXiv:hep-ph/9903535];
%%CITATION = HEP-PH 9903535;%%
A.~Bartl, T.~Gajdosik, E.~Lunghi, A.~Masiero, W.~Porod, H.~Stremnitzer 
and O.~Vives,
%``General flavor blind MSSM and CP violation,''
Phys.\ Rev.\ D {\bf 64}, 076009 (2001)
[arXiv:hep-ph/0103324];
%%CITATION = HEP-PH 0103324;%%
A.~J.~Buras and R.~Fleischer,
%``Bounds on the unitarity triangle, sin(2beta) and K $\to$ pi nu anti-nu  decays in models with minimal flavor violation,''
Phys.\ Rev.\ D {\bf 64}, 115010 (2001)
[arXiv:hep-ph/0104238].
%%CITATION = HEP-PH 0104238;%%

\bibitem{b2rho} M. Convery [BABAR Collaboration], talk presented at the 
meeting of the APS Division of Particles and Fields DPF-2002, Williamsburg, 
Vriginia, 24-28 May, 2002.

\bibitem{db2wilsonSMmW}
A.~J.~Buras, in 'Probing the Standard Model of Particle Interactions', 
F.David and R. Gupta, eds., 1998, Elsevier Science B.V.
%``Weak Hamiltonian, CP violation and rare decays,''
arXiv:hep-ph/9806471;
%%CITATION = HEP-PH 9806471;%%

\bibitem{Buras:1990fn}
A.~J.~Buras, M.~Jamin and P.~H.~Weisz,
%``Leading And Next-To-Leading QCD Corrections To Epsilon Parameter 
%And B0 - Anti-B0 Mixing In The Presence Of A Heavy Top Quark,''
Nucl.\ Phys.\ B {\bf 347}, 491 (1990).
%%CITATION = NUPHA,B347,491;%%

\bibitem{Buchalla:1995vs}
G.~Buchalla, A.~J.~Buras and M.~E.~Lautenbacher,
%``Weak Decays Beyond Leading Logarithms,''
Rev.\ Mod.\ Phys.\  {\bf 68}, 1125 (1996)
[arXiv:hep-ph/9512380].
%%CITATION = HEP-PH 9512380;%%

\bibitem{Becirevic:2001xt}
D.~Becirevic, V.~Gimenez, G.~Martinelli, M.~Papinutto and J.~Reyes,
%``B-parameters of the complete set of matrix elements of Delta(B) = 2  
%operators from the lattice,''
JHEP {\bf 0204}, 025 (2002)
[arXiv:hep-lat/0110091].
%%CITATION = HEP-LAT 0110091;%%

\bibitem{ko2} S. Baek, J. H. Jang, P. Ko and J.H. Park, 
Nucl. Phys. {\bf B 609}, 442 (2001)
[arXiv:hep-ph/0105028].
%%CITATION = HEP-PH 0105028;%%

\bibitem{Buras:1997fb}
A.~J.~Buras and R.~Fleischer,
%``Quark mixing, CP violation and rare decays after the top quark  discovery,''
Adv.\ Ser.\ Direct.\ High Energy Phys.\  {\bf 15}, 65 (1998)
[arXiv:hep-ph/9704376].
%%CITATION = HEP-PH 9704376;%%

\bibitem{Ciuchini:2000de}
M.~Ciuchini {\it et al.},
%``2000 CKM-triangle analysis: A critical review with updated experimental  
%inputs and theoretical parameters,''
JHEP {\bf 0107}, 013 (2001)
[arXiv:hep-ph/0012308].
%%CITATION = HEP-PH 0012308;%%

\bibitem{randall} 
L.~Randall and S.~Su,
%``CP violating lepton asymmetries from B decays and their implication
%for  supersymmetric flavor models,''
Nucl.\ Phys.\ B {\bf 540}, 37 (1999)
[arXiv:hep-ph/9807377];
%%CITATION = HEP-PH 9807377;%%
%\cite{Laplace:2002ik}

\bibitem{ali} A.~Ali, H.~Asatrian and C.~Greub,
%``Inclusive decay rate for B $\to$ X/d + gamma in next-to-leading  
%logarithmic order and CP asymmetry in the standard model,''
Phys.\ Lett.\ B {\bf 429}, 87 (1998) 
[arXiv:hep-ph/9803314].
%%CITATION = HEP-PH 9803314;%%

\bibitem{kn} 
A.~L.~Kagan and M.~Neubert,
%``Direct CP violation in B $\to$ X/s gamma decays as a signature of new  
%physics,''
Phys.\ Rev.\ D {\bf 58}, 094012 (1998)
[arXiv:hep-ph/9803368].
%%CITATION = HEP-PH 9803368;%% 

\bibitem{keum}
A.~G.~Akeroyd, Y.~Y.~Keum and S.~Recksiegel,
%``Effect of supersymmetric phases on the direct CP asymmetry of B 
%$\to$  X/d gamma,''
Phys.\ Lett.\ B {\bf 507}, 252 (2001)
[arXiv:hep-ph/0103008].
%%CITATION = HEP-PH 0103008;%%

\bibitem{buras}
A.~J.~Buras, G.~Colangelo, G.~Isidori, A.~Romanino and L.~Silvestrini,
%``Connections between epsilon'/epsilon and rare kaon decays in
%supersymmetry,''
Nucl.\ Phys.\ B {\bf 566}, 3 (2000)
[arXiv:hep-ph/9908371];
%%CITATION = HEP-PH 9908371;%%
T.~Besmer, C.~Greub and T.~Hurth,
%``Bounds on flavor violating parameters in supersymmetry,''
Nucl.\ Phys.\ B {\bf 609}, 359 (2001)
[arXiv:hep-ph/0105292];
%%CITATION = HEP-PH 0105292;%%

\bibitem{ali02} 
A.~Ali and E.~Lunghi,
%``Implications of B $\to$ rho gamma measurements in the Standard Model 
% and Supersymmetric Theories,''
arXiv:hep-ph/0206242.
%%CITATION = HEP-PH 0206242;%%

\end{thebibliography}
\end{document}